\documentclass[12pt,a4paper]{article}

\usepackage{amssymb,amstext}

\newcommand{\realni}{\ensuremath{\mathbb{R}}}
\newcommand{\grasmanovi}{\ensuremath{\mathbb{G}}}

\newcommand{\diag}{\mathop{\rm diag}\nolimits}
\newcommand{\del}{\partial}
\newcommand{\komut}[2]{\mathop{[#1,#2]}\nolimits}
\newcommand{\LEVAZ}{{\mathop{\langle}\nolimits}}
\newcommand{\DESNAZ}{{\mathop{\rangle}\nolimits}}
\newcommand{\sredvr}[1]{\LEVAZ #1 \DESNAZ}
\newcommand{\ds}{\displaystyle}
\newcommand{\triang}{\triangleright}
\newcommand{\zagrada}[2]{\mathop{\{#1,#2\} }\nolimits}

\newcommand{\cA}{{\cal A}}
\newcommand{\cC}{{\cal C}}
\newcommand{\cD}{{\cal D}}
\newcommand{\cF}{{\cal F}}
\newcommand{\cG}{{\cal G}}
\newcommand{\cH}{{\cal H}}
\newcommand{\cM}{{\cal M}}
\newcommand{\cP}{{\cal P}}
\newcommand{\cS}{{\cal S}}
\newcommand{\cV}{{\cal V}}

\newcommand{\gl}{{\mathfrak{l}}}
\newcommand{\gh}{{\mathfrak{h}}}
\renewcommand{\gg}{{\mathfrak{g}}}
\newcommand{\bd}{{\mathbf{d}}}

\title{Quantum gravity and elementary particles from higher gauge theory}

\author{Tijana Radenkovi\'c$^{1,*}$ and Marko Vojinovi\' c$^{1,\dag}$ \\
\small ${}^{1}$Institute of Physics, University of Belgrade, \\
\small Pregrevica 118, 11080 Belgrade, Serbia \\
\small ${}^*$E-mail: rtijana@ipb.ac.rs $\qquad$ ${}^\dag$E-mail: vmarko@ipb.ac.rs}

\date{}

\begin{document}

\maketitle

\begin{abstract}
We give a brief overview how to couple general relativity to the Standard Model of elementary particles, within the higher gauge theory framework, suitable for the spinfoam quantization procedure. We begin by providing a short review of all relevant mathematical concepts, most notably the idea of a categorical ladder, $3$-groups and generalized parallel transport. Then, we give an explicit construction of the algebraic structure which describes the full Standard Model coupled to Einstein-Cartan gravity, along with the classical action, written in the form suitable for the spinfoam quantization procedure. We emphasize the usefulness of the $3$-group concept as a superior tool to describe gauge symmetry, compared to an ordinary Lie group, as well as the possibility to employ this new structure to classify matter fields and study their spectrum, including the origin of fermion families.
\end{abstract}

\section{Introduction}

The quantization of the gravitational field is one of the most fundamental open problems of modern theoretical physics. Since the inceptions of general relativity (GR) and quantum field theory (QFT), many attempts have been made over the years to unify the two into a self-consistent description of gravitational and matter fields as basic building blocks of nature. Some of the attempts have developed into vast research areas, such as String Theory, Loop Quantum Gravity, Causal Set Theory, and so on. One of the prominent approaches is Loop Quantum Gravity (LQG) \cite{Rovelli,Thiemann}, which has branched into the canonical and covariant frameworks, the latter known as the {\em spinfoam} approach \cite{RovelliVidotto}.

The spinfoam approach to the quantization of the gravitational field revolves around the idea of providing a precise mathematical definition to the Feynman path integral for the gravitational field,
$$
Z = \int \cD g \; e^{iS_{GR}[g]}\,,
$$
where $g$ denotes the gravitational degrees of freedom, and $S_{GR}[g]$ is the GR action expressed in terms of variables $g$. The strategy of defining the path integral can be roughly expressed in three main steps, called the {\em spinfoam quantization procedure}:
\begin{enumerate}
\item Choose convenient variables $g$ and rewrite the classical action in the form
\begin{equation} \label{OpsteDejstvo}
S_{GR}[g] = S_{\text{topological}}[g] + S_{\text{simp}}[g]\,,
\end{equation}
where the first term represents a topological theory (with no propagating degrees of freedom), while the second term corresponds to the so-called {\em simplicity constraint} terms, whose purpose is to transform the full action into a realistic non-topological action with propagating degrees of freedom.
\item Employ the methods of topological quantum field theory (TQFT) to define the path integral for the topological part of the action. This is typically implemented by passing from a smooth spacetime manifold to a simplicial complex (triangulation), and writing the path integral in the form of a discrete state sum,
$$
  Z = \sum_g \prod_v \cA_v(g) \prod_{\epsilon} \cA_{\epsilon}(g) \prod_{\Delta} \cA_{\Delta}(g) \prod_{\tau} \cA_{\tau}(g) \prod_{\sigma} \cA_{\sigma}(g)\,.
$$
Here $g$ represents the gravitational field variables living on the vertices $v$, edges $\epsilon$, triangles $\Delta$, terahedra $\tau$, and $4$-simplices $\sigma$ of the simplicial complex, describing its geometry, while the corresponding amplitudes $\cA_v(g)$, \dots, $\cA_{\sigma}(g)$ are chosen to render the whole state sum $Z$ independent of the particular choice of the triangulation of the spacetime manifold.
\item Enforce the simplicity constraints of the theory by a suitable deformation of the amplitudes $\cA$ and the set of independent variables $g$, thereby obtaining a modified state sum $Z$ which corresponds to one possible rigorous definition of the realistic gravitational path integral.
\end{enumerate}

Since its inception, the spinfoam quantization procedure has been formulated and implemented for various choices of the classical action, leading to a plethora of {\em spinfoam models} of quantum gravity, starting from the Ponzano-Regge model for $3D$ gravity \cite{PonzanoRegge}, and leading up to the currently most sophisticated EPRL/FK model for the realistic $4D$ case \cite{EPRL,FK}. However, one property common to all spinfoam models is the fact that they all describe pure gravity, without matter fields. This is due to the common choice of the classical action --- it is the well known $BF$ theory \cite{BFgravity}, which is usually defined for the Lorentz group $SO(3,1)$, with some form of the simplicity constraint terms. The prototype description of GR in this form is the Plebanski action \cite{Plebanski}. The reason why matter fields are absent from all such models lies in the fact that the $BF$ action does not feature tetrad fields at the fundamental level. Instead, the tetrads appear as a consequence of classical equations of motion, and are thus inherently classical, on-shell quantities. This renders the approach based on the $BF$ theory incapable of adding matter fields at the quantum level, since matter is coupled to gravity using precisely the tetrad fields.

The issue of the absence of the tetrad fields at the fundamental level has been successfully resolved in \cite{MV}, where a categorical generalization has been made, and the $2BF$ action (introduced in \cite{GirelliPfeifferPopescu,FariaMartinsMikovic}) has been employed to build an action for GR, featuring tetrads explicitly in the topological sector of the action. The categorical generalization is based on a concept of a {\em categorical ladder}, an abstraction scheme introducing a chain of new objects: from categories to $2$-categories to $3$-categories and so forth. This powerful mathematical language gave rise to the idea that the notion of gauge symmetry in physics may be described by objects other than Lie groups. The new approach is called {\em higher gauge theory} (HGT), see \cite{BaezHuerta} for an introduction. In the context of the spinfoam quantization procedure, HGT has been successfully applied to build a quantum gravity model, based on the Poincar\'e $2$-group \cite{CraneSheppeard} as a gauge symmetry structure, and the corresponding $2BF$ action, leading to the so-called {\em spincube model} of quantum gravity \cite{MV}. Having the tetrads as fundamental fields in the $2BF$ action, the new model could be extended to include matter fields in a straightforward way. Nevertheless, the matter field action does not have the form analogous to (\ref{OpsteDejstvo}), which renders the steps 2 and 3 of the spinfoam quantization procedure moot, since they can be applied only to the gravitational sector of the theory.

Thus, a natural need appeared to generalize the theory once more, in order to include the matter fields into the topological sector of the theory, in a similar way that was done to include the tetrad fields. The basic idea was to pass from the notion of a $2$-group to a notion of a $3$-group as a mathematical descriptor of gauge symmetry \cite{BaezHuerta,MartinsPicken,Wang}, giving rise to a topological $3BF$ action. With suitable simplicity constraint terms added, a $3BF$ action perfectly fits together all fields necessary for a unified description of quantum gravity coupled to matter fields --- it features tetrads, spin connection, gauge fields, scalar fields and fermions. The explicit construction was done in \cite{JHEP}, where the full Standard Model (SM) coupled to GR in the Einstein-Cartan formulation was rewritten in the form (\ref{OpsteDejstvo}), suitable for the implementation of the spinfoam quantization procedure and building a full quantum theory. This demostrates the power and expressiveness of the HGT approach, and it provides us with novel mathematical tools to study the algebraic properties of the matter sector of the SM, in analogy to the gauge field sector which is being described in terms of ordinary Lie groups. In this paper we will review the essential properties of the new approach.

The layout of the paper is the following. In section \ref{SecII} we give a brief introduction to the category theory, categorical ladder, and the notion of $n$-groups. Our attention focuses on $3$-groups, in particular their representation in terms of $2$-crossed modules. Section \ref{SecIII} reviews the construction and general properties of the $3BF$ action, and its relationship with the $3$-group structure. Then, in section \ref{SecIV} we apply this developed formalism to construct the {\em Standard Model $3$-group}, and explicitly build the action for the Standard Model coupled to Einstein-Cartan gravity in the form of the $3BF$ action with suitable simplicity constraints. Section \ref{SecV} contains our concluding remarks.

\section{\label{SecII}Category theory and $3$-groups}

Let us begin by giving a short introduction to the category theory, and in particular the notion of {\em category theory ladder}, a concept used in higher gauge theory to generalize the notion of gauge symmetry. A nice introduction to this topic can be found in \cite{BaezHuerta} and further technical details in \cite{MartinsPicken,Wang}.

A category $\cC = (Obj, Mor )$ is a structure which has objects and morphisms between them,
$$
X,Y,Z,\dots \in Obj\,, \qquad f,g,h,\dots\in Mor\,,
$$
where
$$
f:X\to Y, \quad  g:Z\to X, \quad h:X\to Y, \dots
$$
such that certain rules are respected, like the associativity of composition of morphisms, and similar. Similarly, a $2$-category $\cC_2 = (Obj, Mor_1, Mor_2 )$ is a structure which has objects, morphisms between them, and morphisms between morphisms, called $2$-morphisms,
$$
X,Y,Z,\dots \in Obj\,, \qquad f,g,h,\dots\in Mor_1\,, \qquad \alpha,\beta,\dots \in Mor_2\,,
$$
where
$$
f:X\to Y, \quad  g:Z\to X, \quad h:X\to Y, \dots \qquad \alpha:f\to h\,, \dots
$$
such that similar rules about compositions are respected. Then, a $3$-category $\cC_3 = (Obj, Mor_1, Mor_2, Mor_3 )$ additionally has morphisms between $2$-mor\-ph\-isms, called $3$-morphisms,
$$
\Theta,\Phi,\dots \in Mor_3\,, \qquad \Theta:\alpha\to \beta\,, \dots
$$
again with a certain set of axioms about compositions of various $n$-mor\-ph\-isms. One can further generalize these structures to introduce $4$-categories, $n$-categories, $\infty$-categories, etc. The process of raising the ``dimensionality'' of a categorical structure is called a {\em categorical ladder}.

It is useful to understand other algebraic structures as special cases of categories. As a particularly important example, the algebraic structure of a {\em group} is a special case of a category --- it is a category with only one object, while all morphisms (i.e., group elements) are invertible. It is straightforward to verify that axioms of a group follow from this definition and the axioms of a category. Any group can be represented in this way, for example finite groups, Lie groups, and so on.

The notion of a categorical ladder then provides us with a natural way to introduce novel, more general algebraic structures, by extending the above definition to $2$-categories, $3$-categories, etc. In particular,
\begin{itemize}
\item a $2$-group is a $2$-category with only one object, while all $1$-morphisms and $2$-morphisms are invertible;
\item a $3$-group is a $3$-category with only one object, while all $1$-morphisms, $2$-morphisms and $3$-morphisms are invertible.
\end{itemize}
It is important to emphasize that an $n$-group is not a particular type of group. Instead, it is a different algebraic structure, which shares some of the features of groups, but is governed by a qualitatively different set of axioms.

The framework of higher gauge theory is centered around the idea that gauge symmetries in physics can be better described using these alternative algebraic structures than using the ordinary Lie groups. To that end, our attention will mostly focus on the so-called Lie $3$-groups and their corresponding Lie $3$-algebras. While the abstract definition in terms of $n$-category theory is particularly appealing from the conceptual point of view, for applications in physics there exists a more practical way to talk about $3$-group. Namely, every strict Lie $3$-group is known to be equivalent to a so-called {\em $2$-crossed module}, defined as an exact sequence of three Lie groups $G$, $H$ and $L$,
\begin{equation} \label{DvaUkrsteniModul}
L \stackrel{\delta}{\to} H \stackrel{\del}{\to} G\,,
\end{equation}
and equipped with two ``boundary homomorphisms'' $\delta$ and $\del$, an action $\triang$ of $G$ onto $G$, $H$ and $L$,
$$
\triang: G\times G \to G\,, \qquad
\triang: G\times H \to H\,, \qquad
\triang: G\times L \to L\,,
$$
and a bracket operation called {\em Peiffer lifting} over $H$ to $L$,
$$
\zagrada{\_\;}{\_} : H\times H \to L\,.
$$
Certain set of axioms is assumed to hold true among all these maps. In particular, for all $g\in G$, $h\in H$ and $l\in L$, we have:
\begin{itemize}
\item the axiom stating that (\ref{DvaUkrsteniModul}) is an exact sequence,
\begin{equation} \label{axjedan}
\del \delta = 1_G\,,
\end{equation}
\item the axiom specifying that the action of $G$ onto itself is conjugation,
\begin{equation} \label{axdva}
g \triang g_0 = g\, g_0\, g^{-1}\,,
\end{equation}
\item the axioms stating that the action of $G$ on $H$ and $L$ is equivariant with respect to homomorphisms $\del$ and $\delta$ and the Peiffer lifting,
\begin{equation} \label{axtri}
\begin{array}{lcl}
g \triang \del h & = & \del (g\triang h)\,, \\
g \triang \delta l & = & \delta (g\triang l)\,, \\
g \triang \zagrada{h_1}{h_2} & = & \zagrada{g\triang h_1}{g\triang h_2}\,, \\
\end{array}
\end{equation}
\item and finally the axioms determining the properties of the Peiffer lifting,
\begin{equation} \label{axcetiri}
\begin{array}{lcl}
  \delta \zagrada{h_1}{h_2} & = & h_1 h_2 h_1^{-1} (\del h_1) \triang h_2^{-1}\,, \\
  \zagrada{\delta l_1}{ \delta l_2} & = & l_1 l_2 l_1^{-1} l_2^{-1}\,, \\
  \zagrada{h_1 h_2}{ h_3} & = & \zagrada{h_1}{ h_2 h_3 h_2^{-1}} \del h_1 \triang \zagrada{h_2}{h_3}\,, \\
  \zagrada{\delta l}{ h} \zagrada{h}{\delta l} & = & l (\del h \triang l^{-1})\,. \\
\end{array}
\end{equation}
\end{itemize}

Since it is constructed from three Lie groups, a Lie $3$-group has a corresponding Lie $3$-algebra, also called a {\em differential $2$-crossed module},
$$
\gl \stackrel{\delta}{\to} \gh \stackrel{\del}{\to} \gg\,,
$$
where $\gl$, $\gh$, $\gg$ are Lie algebras of $L$, $H$, $G$, the maps $\delta$, $\del$, $\triang$ and $\zagrada{\_\;}{\_}$ are inherited from the $3$-group via natural linearization, and finally, the set of corresponding axioms applies. In addition to all this, Lie algebras have their own usual Lie structure --- the generators,
$$
T_A \in \gl\,, \qquad t_a \in \gh\,, \qquad \tau_{\alpha} \in \gg\,
$$
the corresponding structure constants,
$$
\komut{T_A}{T_B} = f_{AB}{}^C T_C\,, \qquad
\komut{t_a}{t_b} = f_{ab}{}^c t_c\,, \qquad
\komut{\tau_{\alpha}}{\tau_{\beta}} = f_{\alpha\beta}{}^{\gamma} \tau_{\gamma}\,,
$$
and $G$-invariant nondegenerate symmetric bilinear forms (for example Killing forms),
$$
\sredvr{T_A,T_B}_{\gl} = g_{AB}\,, \qquad
\sredvr{t_a,t_b}_{\gh} = g_{ab}\,, \qquad
\sredvr{\tau_{\alpha},\tau_{\beta}}_{\gg} = g_{\alpha\beta} \,.
$$

The main purpose of the $3$-group structure is to {\em generalize the notion of parallel transport} from curves to surfaces to volumes. Namely, given a $4$-dimensional manifold $\cM$, one defines a $3$-connection $(\alpha,\beta,\gamma)$ as a triple of $3$-algebra-valued differential forms,
$$
\begin{array}{lcll}
\alpha & = & \ds \vphantom{\frac{1}{2}} \alpha^{\alpha}{}_{\mu}(x) \, \tau_{\alpha}\, \bd x^{\mu} & \in \Lambda^1(\cM,\gg) \,, \vphantom{\ds\int} \\
\beta & = & \ds\frac{1}{2} \beta^a{}_{\mu\nu}(x) \, t_a \, \bd x^{\mu} \wedge \bd x^{\nu} & \in \Lambda^2(\cM,\gh) \,, \vphantom{\ds\int^A} \\
\gamma & = & \ds\frac{1}{3!} \gamma^A{}_{\mu\nu\rho}(x) \, T_A \, \bd x^{\mu} \wedge \bd x^{\nu} \wedge \bd x^{\rho} & \in \Lambda^3(\cM,\gl)\,. \vphantom{\ds\int^A} \\
\end{array}
$$
Then one can introduce the line, surface and volume holonomies,
$$
g = \cP\! \exp \int_{\cP_1} \alpha\,, \qquad
h = \cS\!\exp \int_{\cS_2} \beta\,, \qquad
l = \cV\!\exp \int_{\cV_3} \gamma\,,
$$
and corresponding curvature forms,
$$
\begin{array}{lcl}
\cF & = & \bd \alpha + \alpha \wedge \alpha - \del \beta \,, \\
\cG & = & \bd \beta + \alpha \wedge^{\triang} \beta - \delta \gamma \,, \\
\cH & = & \bd \gamma + \alpha \wedge^{\triang} \gamma - \{ \beta \wedge \beta \} \,. \\
\end{array}
$$
The $3$-group structure ensures that all these quantities are well defined, in particular the surface- and volume-ordered exponentials and the respective holonomies.

\section{\label{SecIII}Higher gauge theories}

The basic idea behind the higher gauge theory approach is to employ the structure of $n$-groups as a mathematical representation of gauge symmetries in physics, generalizing the ordinary notion of gauge symmetry described via a Lie group. Namely, in ordinary gauge theory, the prototype action functional was the so-called $BF$ action \cite{BFgravity}, based on a chosen gauge group $G$. In the HGT approach, one generalizes the $BF$ action in accord with the chosen $n$-group structure, leading to the $nBF$ action. For the case of $3$-groups, one defines a $3BF$ action as:
$$
S_{3BF} = \int_{\cM} \sredvr{B \wedge \cF }_{\gg} + \sredvr{ C\wedge \cG}_{\gh} + \sredvr{ D \wedge \cH}_{\gl}\,.
$$
Here $B$, $C$, and $D$ are Lagrange multipliers, in particular a $\gg$-valued $2$-form, an $\gh$-valued $1$-form, and an $\gl$-valued $0$-form, respectively.

As in the case of a $BF$ theory, one can demonstrate that $3BF$ theory is a topological gauge theory, having no local propagating degrees of freedom. Nevertheless, it can be transformed into a physically relevant action by adding the so-called {\em simplicity constraint terms} to the action, changing the dynamical structure of the theory. The prototype of this procedure is represented by transforming the topological $BF$ theory based on the Lorentz group $SO(3,1)$ into a Plebanski action \cite{Plebanski}, which describes general relativity.

One can even do more, and provide a physical interpretation of the Lagrange multipliers $C$ and $D$ in the $3BF$ action, as follows:
\begin{itemize}
\item the $\gh$-valued $1$-form $C$ can be interpreted as the tetrad field, if $H=\realni^4$ is the spacetime translation group,
$$
C \to e = e^a{}_{\mu}(x) \, t_a \, \bd x^{\mu}\,,
$$
\item the $\gl$-valued $0$-form $D$ can be interpreted as the set of real-valued matter fields, given some Lie group $L$,
$$
D \to \phi = \phi^A(x) \, T_A\,.
$$
\end{itemize}
An interested reader can see \cite{JHEP} for further details.

\section{\label{SecIV}The Standard Model}

One natural question that can be asked is what choice of a $3$-group can be relevant for physics. There are various answers to this question, but perhaps the most illustrative example is a choice of the $3$-group which reproduces the Standard Model of elementary particles, coupled to general relativity in the Einstein-Cartan version. This is called the {\em Standard Model $3$-group}, and in the remainder of this section we will demonstrate how it can be constructed, step by step.

The first step is to specify the groups $G$ and $H$ as the usual Lorentz, internal, and translational symmetries:
$$
G = SO(3,1) \times SU(3) \times SU(2) \times U(1)\,, \qquad H = \realni^4\,.
$$
Note that the Poincar\'e group has been broken into the separate Lorentz and translational parts, and these have been associated with two different groups within the $3$-group structure.

The next step is to define the homomorphisms $\delta$ and $\del$, as well as the Peiffer lifting, to be trivial,
$$
\delta l = 1_H = 0\,, \qquad \del \vec{v} = 1_G\,,
$$
and
$$
\zagrada{\vec{u}}{\vec{v}} = 1_L\,,
$$
for all $l\in L$ and $\vec{u},\vec{v} \in H$. Additionally, we define the action of the group $G$ on $H$ via vector representation for the $SO(3,1)$ sector and via trivial representation for the $SU(3)\times SU(2) \times U(1)$ sector. Finally, the choice of the group $L$ and the action of $G$ on $L$ will be discussed below. But already now one can verify that all axioms (\ref{axjedan})--(\ref{axcetiri}) are satisfied, thus making sure that these choices represent one genuine $3$-group.

The next step is to choose the group $L$. One general property of $L$ that can be determined immediately comes from the second axiom in (\ref{axcetiri}). Namely, due to the trivial choices for the Peiffer lifting and the homomorphism $\delta$, the axiom implies that $L$ must be Abelian. Aside from this, the choice of the group $L$ is guided by physical requirements, as follows.

Begin by rewriting the $3BF$ action in the form
$$
S_{3BF} = \int_{\cM} B^{\alpha} \wedge \cF^{\beta} g_{\alpha\beta} + e^a \wedge \cG^b g_{ab} + \phi^A \cH^B g_{AB}\,.
$$
Since the group $G$ is a direct product of the Lorentz and internal groups, the corresponding indices $\alpha$ of $G$ split according to this structure, as $\alpha = ( ab\,,\,i)$, leading to the corresponding splitting of the connection $\alpha$ and its curvature $\cF$,
$$
\alpha = \omega^{ab} J_{ab} + A^i \tau_i\,, \qquad \cF = R^{ab} J_{ab} + F^i \tau_i\,.
$$
Here $\omega^{ab}$ is the ordinary spin connection $1$-form, $J_{ab}$ are Lorentz generators, while $A^i$ are internal gauge potential $1$-forms and $\tau_i$ the generators of $SU(3)\times SU(2) \times U(1)$. Also, $R^{ab}$ and $F^i$ are the Riemann curvature and gauge field strength $2$-forms, respectively. Also, given that the action of $SO(3,1)$ onto $H = \realni^4$ is via vector representation, and given that the bilinear symmetric nondegenerate form for $H$ must be $G$-invariant, the only available choice is
$$
g_{ab} = \eta_{ab} \equiv \diag (-1,+1,+1,+1)\,.
$$
Finally, given that the matter fields are elements in the Lie algebra $\gl$ of the group $L$, namely $\phi = \phi^A T_A$, we observe that there should be precisely one real-valued field $\phi^A(x)$ for each generator $T_A \in \gl$. This information allows us to determine the dimension of the algebra $\gl$, by counting the total number of real-valued components of all matter fields in the Standard Model. The matter fields have two sectors --- fermions and the Higgs.

The number of the real-valued components of all fermion fields can be counted according to the following scheme:
$$
\left.
\begin{array}{cccc}
\ds \left( \nu_e \atop e^-  \right)_L & \ds \left( u_r \atop d_r \right)_L & \ds \left( u_g \atop d_g \right)_L & \ds \left( u_b \atop d_b \right)_L \\
 & & & \\
 (\nu_e)_R & (u_r)_R & (u_g)_R & (u_b)_R \\
 & & & \\
 (e^-)_R & (d_r)_R & (d_g)_R & (d_b)_R \\
\end{array}
\right\} = 16\;\; \frac{\text{Weyl spinors}}{\text{family}} \times
$$
$$
\times 3 \text{ families } \times 4 \;\; \frac{\text{real-valued fields}}{\text{Weyl spinor}} \;\; = 192 \text{ real-valued fields } \phi^A\,.
$$
Similarly, the Higgs sector gives us:
$$
\left. \left( \phi^+ \atop \phi_0 \right) \right\} = 2 \text{ complex scalar fields } = 4 \text{ real-valued fields } \phi^A\,.
$$
This suggests the structure for $L$ in the form:
$$
L = L_{\text{fermion}} \times L_{\text{Higgs}}\,, \qquad \dim L_{\text{fermion}} = 192\,, \qquad \dim L_{\text{Higgs}} = 4\,.
$$

The structure of $L$ can be further understood by looking at the action of the gauge group $G$ on various components of fields $\phi^A$. This is fixed by the choice of the action of $G$ on $L$, chosen as follows. Given that $G$ is constructed from Lorentz and internal gauge symmetry groups, the action $\triang:G\times L \to L$ specifies the transformation properties of each real-valued field $\phi^A$ with respect to those symmetries. For example, if we look at a Weyl spinor $u_b$ that sits in the doublet
$$
 \left( u_b \atop d_b \right)_L \,,
$$
the action $g\triang u_b$ (where $g\in SO(3,1)\times SU(3) \times SU(2) \times U(1)$) encodes that $u_b$ consists of $4$ real-valued fields which transform as:
\begin{itemize}
\item a left-handed spinor with respect to $SO(3,1)$,
\item as a ``blue'' component of the fundamental representation of $SU(3)$,
\item and as ``isospin $+\frac{1}{2}$'' of the left doublet with respect to $SU(2)\times U(1)$.
\end{itemize}
The action $\triang: G\times L \to L$ similarly defines the transformation properties for all other fermions in the theory, as well as for the Higgs field.

From such a definition of the action $\triang$, one can observe that $G$ acts on $L$ in precisely the same way across the three fermion families. This implies that $L_{\text{fermion}}$ can be written as
$$
L_{\text{fermion}} = L_{\text{1st family}} \times L_{\text{2nd family}} \times L_{\text{3rd family}}\,, \qquad \dim L_{k\text{-th family}} = 64\,.
$$

Ultimately, given that the components of Weys spinors mutually anticommute, given that the group $L$ is Abelian, and given that it has the structure and dimension as given above, we can fix the choice of the group $L$ which corresponds to the Standard Model as
$$
L= \mathbb{R}^4(\mathbb{C}) \times \mathbb{R}^{64}(\grasmanovi) \times \mathbb{R}^{64}(\grasmanovi) \times \mathbb{R}^{64}(\grasmanovi)\,,
$$
where $\grasmanovi$ is the algebra of Grassmann numbers. This completes the construction of the Standard Model $3$-group.

The final step in specifying the theory is to spell out its classical action. As was previously discussed, the action has the form of a $3BF$ action, with the addition of appropriate simplicity constraints which will transform it into a non-topological theory, i.e., a theory with local propagating degrees of freedom. The choice of the Standard Model $3$-group completely fixes the structure of the $3BF$ action, and the only thing left to do is to add the appropriate simplicity constraints. The details of the construction of these terms is given in detail in \cite{JHEP}, and will not be repeated here. We will only quote the result,
$$
S_{SM+EC} = S_{3BF} + S_{\text{simp}}\,,
$$
where
$$
S_{3BF} =\int B_{\hat{\alpha}}\wedge \mathcal{F}^{\hat{\alpha}}+e_{\hat {a}}\wedge \mathcal{G}^{\hat{a}}+\phi_{\hat{A}}\wedge \mathcal{H}^{\hat{A}}\vphantom{\ds\int}\,,
$$
and
$$
S_{\text{simp}} = \left(B_{\hat{\alpha}}-C_{\hat{\alpha}}{}^{\hat{\beta}}M_{cd\hat{\beta}} e^c\wedge e^d\right)\wedge\lambda^{\hat{\alpha}} -\left(\gamma_{\hat{A}}-e^a\wedge e^b \wedge e^c C_{\hat{A}}{}^{\hat{B}}M_{abc\hat{B}}\right)\wedge{\lambda}^{\hat{A}}\vphantom{\ds\int}
$$
$$
-4\pi i\, l_p^2\,\varepsilon_{abcd}{e^a\wedge e^b \wedge \beta^c \phi_{\hat{A}}T^{d}{}^{\hat{A}}{}_{\hat{B}}\phi{}^{\hat{B}}} \vphantom{\ds\int} 
$$
$$
+{\zeta^{ab}{}_{\hat{\alpha}}}\wedge\left({M{}_{ab}{}^{\hat{\alpha}}}\varepsilon^{cdef}e_c\wedge e_d \wedge e_e \wedge e_f - F^{\hat{\alpha}} \wedge e_c \wedge e_d\right) \vphantom{\ds\int} 
$$
$$
+{\zeta^{ab}}{}_{\hat{A}}\wedge\left({M_{abc}}{}^{\hat{A}}\varepsilon^{cdef}e_d\wedge e_e \wedge e_f- F^{\hat{A}} \wedge e_a \wedge e_b\right)\vphantom{\ds\int} 
$$
$$
- \varepsilon_{abcd}e^a\wedge e^b \wedge e^c \wedge e^d \; \left( \Lambda + M_{\hat{A}\hat{B}} \phi^{\hat{A}}\phi^{\hat{B}} +Y_{\hat{A}\hat{B}\hat{C}} \phi^{\hat{A}} \phi{}^{\hat{B}} \phi{}^{\hat{C}} + L_{\hat{A}\hat{B}\hat{C}\hat{D}} \phi^{\hat{A}} \phi{}^{\hat{B}} \phi{}^{\hat{C}} \phi{}^{\hat{D}}\right)\,. \vphantom{\ds\int} 
$$
See \cite{JHEP} for details and notation.

By varying the action with respect to all variables, and with a little technical effort, one can demonstrate that the corresponding equations of motion are precisely the classical equations of the Standard Model, coupled to general relativity in the Einstein-Cartan formulation.

\section{\label{SecV}Conclusions}

Let us summarize the results of the paper. In section \ref{SecII} we have given a short introduction into the category theory, introduced the notions of categorical ladder and $n$-categories, and in the resulting framework, provided a definition for the notion of an $n$-group. Our attention focused on the case of $3$-groups, which are relevant for applications in physics, and the equivalent notion of a $2$-crossed module, which is more convenient for practical applications. Section \ref{SecIII} was devoted to introducing the higher gauge theory formalism and the $3BF$ action corresponding to a choice of a $3$-group, as a generalization of the well known $BF$ action in terms of the categorical ladder. Also, we have interpreted the additional Lagrange multipliers appearing in the $3BF$ action as the tetrad and matter fields, providing the setup for the application in physics. This application was then demonstrated in detail in section \ref{SecIV}, where the Standard Model 3-group has been defined, and utilized to construct a physically relevant constrained $3BF$ action, which is classically equivalent to the Standard Model of elementary particles coupled to general relativity in the Einstein-Cartan formulation. This is the main result, which successfully establishes the first step of the spinfoam quantization procedure, and opens up a possibility of straightforward implementation of the second and third steps, hopefully leading to a full model of quantum gravity with matter.

It should be noted that the most important feature of the higher gauge theory framework is its ability to treat gravity, gauge fields, fermions and scalar fields on completely equal footing, describing all of them via the underlying algebraic structure of a $3$-group. The $3$-group also provides us with a natural geometric description of a generalized notion of parallel transport, namely along a surface and along a volume, in addition to the standard notion of parallel transport along a curve. This relationship opens up a possibility for a fully geometric interpretation of all fields present in physics.

Moreover, just as the gauge group dictates the number and properties of gauge fields in Yang-Mills theories, the sector of the $3$-group described by the Lie group $L$ determines the number and properties of the fermion and scalar fields. This fact enables us to classify the spectrum of matter fields in terms of group theory, generalizing the constructions present in the Standard Model, where only gauge fields are classified in such terms. The choice of the group $L$ thus opens up novel avenues for research on the unification of all fields, and specifically the origin of particle families, Higgs and fermion sectors, and so on.

Finally, the higher gauge theory framework may have applications in other areas of physics and mathematics as well, and various possible research directions are yet to be explored.

\bigskip

\textbf{Acknowledgments.} The authors have been supported by the Ministry of Education, Science and Technological Development of the Republic of Serbia.


\begin{thebibliography}{99}

\bibitem{Rovelli}
C. Rovelli,
{\it Quantum Gravity}, 
Cambridge University Press, Cambridge (2004).

\bibitem{Thiemann}
T. Thiemann,
{\it Modern Canonical Quantum General Relativity},
Cambridge University Press, Cambridge (2007).

\bibitem{RovelliVidotto}
C. Rovelli and F. Vidotto,
{\it Covariant Loop Quantum Gravity},  
Cambridge University Press, Cambridge (2014).

\bibitem{PonzanoRegge}
G. Ponzano and T. Regge, 
%{\it Semiclassical limit of Racah coefficients}, 
{\it Spectroscopic and Group Theoretical Methods in Physics}, edited by F. Block, North Holland, Amsterdam (1968).

\bibitem{EPRL}
J. Engle, E. R. Livine, R. Pereira and C. Rovelli,
%{\it LQG vertex with finite Immirzi parameter},
{\it Nucl. Phys.} {\bf B799}, 136 (2008),
\texttt{arXiv:0711.0146}.

\bibitem{FK}
L. Freidel and K. Krasnov,
%{\it A New Spin Foam Model for 4d Gravity},
{\it Class. Quant. Grav.} {\bf 25}, 125018 (2008),
\texttt{arXiv:0708.1595}.

\bibitem{BFgravity}
M. Celada, D. Gonz\' alez and M. Montesinos,
%{\it BF gravity},
{\it Class. Quant. Grav.} {\bf 33}, 213001 {(2016)},
{\tt arXiv:1610.02020}.

\bibitem{Plebanski}
J. F. Plebanski, 
%{\it On the separation of Einsteinian substructures},
{\it J. Math. Phys.} {\bf 18}, 2511 (1977).

\bibitem{MV}
A. Mikovi\'c and M. Vojinovi\'c,
%{\it Poincar\'e 2-group and quantum gravity},
{\it Class. Quant. Grav.} {\bf 29}, 165003 (2012),
{\tt arXiv:1110.4694}.

\bibitem{GirelliPfeifferPopescu}
F. Girelli, H. Pfeiffer and E. M. Popescu,
%{\it Topological Higher Gauge Theory - from BF to BFCG theory},
{\it Jour. Math. Phys.} {\bf 49}, 032503 (2008),
{\tt arXiv:0708.3051}.

\bibitem{FariaMartinsMikovic}
J. F. Martins and A. Mikovi\'c,
%{\it Lie crossed modules and gauge-invariant actions for 2-BF theories},
{\it Adv. Theor. Math. Phys.} {\bf 15}, 1059 (2011),
{\tt arXiv:1006.0903}.

\bibitem{BaezHuerta}
J. C. Baez and J. Huerta,
%{\it An Invitation to Higher Gauge Theory},
{\it Gen. Relativ. Gravit.} {\bf 43}, 2335 (2011),
{\tt arXiv:1003.4485}.

\bibitem{CraneSheppeard}
L. Crane and M. D. Sheppeard, 
%{\it 2-categorical Poincare Representations and State Sum Applications},
{\tt arXiv:math/0306440.}

\bibitem{MartinsPicken}
J. F. Martins and R. Picken,
%{\it The fundamental Gray 3-groupoid of a smooth manifold and local 3-dimensional holonomy based on a 2-crossed module},
{\it Differ. Geom. Appl. Jour.} {\bf 29}, 179 {(2011),}
{\tt arXiv:0907.2566.} 

\bibitem{Wang}
W. Wang,
%{\it On 3-gauge transformations, 3-curvature and Gray-categories},
{\it Jour. Math. Phys.} {\bf 55}, 043506 (2014),
\texttt{arXiv:1311.3796.}

\bibitem{JHEP}
T. Radenkovi\'c and M. Vojinovi\'c,
%{\it Higher Gauge Theories Based on 3-groups},
{\it JHEP} {\bf 10}, 222 (2019),
{\tt arXiv:1904.07566}.

\end{thebibliography}
\end{document}